\documentclass[aps,twocolumn,pra,superscriptaddress,groupedaddress]{revtex4}

\usepackage{amssymb} \usepackage{color,graphicx} \usepackage{amsmath}
\usepackage{amsbsy} \usepackage{amsthm}
\usepackage{svg}
\usepackage{float} \usepackage{braket}
\usepackage{placeins}
\usepackage[colorlinks=true,citecolor=blue,linkcolor=red,urlcolor=red]{hyperref}
\usepackage[sort&compress]{natbib}
\usepackage{comment}
\usepackage{stackengine}
\usepackage{soul}
\usepackage{siunitx}

\newcommand{\bq}{\begin{equation}} \newcommand{\eq}{\end{equation}}
\newcommand{\bqali}{\bq\begin{aligned}}
\newcommand{\eqali}{\end{aligned}\eq}
\newcommand\D{\operatorname{d}\!}

\newcommand\kb{k_\text{\tiny B}}

\newcommand\rC{r_\text{\tiny C}}

\newcommand{\matteo}[1]{{\color{black}#1}}
\newcommand{\ario}[1]{{\color{black}#1}}
\newcommand{\sandro}[1]{{\color{black}#1}}

\begin{document}

\author{Davide Giordano Ario Altamura}
\email{davidegiordanoario.altamura@phd.units.it}
\affiliation{Department of Physics, University of Trieste, Strada Costiera 11, 34151 Trieste, Italy}
\affiliation{Istituto Nazionale di Fisica Nucleare, Trieste Section, Via Valerio 2, 34127 Trieste, Italy}

\author{Matteo Carlesso}
\affiliation{Department of Physics, University of Trieste, Strada Costiera 11, 34151 Trieste, Italy}
\affiliation{Istituto Nazionale di Fisica Nucleare, Trieste Section, Via Valerio 2, 34127 Trieste, Italy}
\affiliation{Centre for Quantum Materials and Technologies,
School of Mathematics and Physics, Queens University, Belfast BT7 1NN, United Kingdom}

\author{Sandro Donadi}
\email{s.donadi@qub.ac.uk}
\affiliation{Centre for Quantum Materials and Technologies,
School of Mathematics and Physics, Queens University, Belfast BT7 1NN, United Kingdom}

\author{Angelo Bassi}
\affiliation{Department of Physics, University of Trieste, Strada Costiera 11, 34151 Trieste, Italy}
\affiliation{Istituto Nazionale di Fisica Nucleare, Trieste Section, Via Valerio 2, 34127 Trieste, Italy}

\title{Non-interferometric rotational test of the Continuous Spontaneous Localisation model: enhancement of the collapse noise  through shape optimisation}

\date{\today}
\begin{abstract}  
The Continuous Spontaneous Localisation (CSL) model is the most studied among collapse models, which  describes the breakdown of the superposition principle for macroscopic systems.  Here, we derive an upper bound on the parameters of the model by applying it to the rotational noise measured in a recent short-distance gravity experiment [Lee \textit{et al}., Phys. Rev. Lett. \textbf{124}, 101101 (2020)]. Specifically, considering the noise affecting the rotational motion, we found that despite being a table-top experiment the bound is only one order of magnitude weaker than that from LIGO for the relevant values of the collapse parameter. Further, we analyse possible ways to optimise the shape of the test mass to enhance the collapse noise by several orders of magnitude and eventually derive stronger bounds that can address the unexplored region of the CSL parameters space.
\end{abstract}
\maketitle

\section{Introduction}
The quantum-to-classical transition is still an open issue in quantum physics. On top of the theoretical and conceptual problems, assessing if and where the transition occurs is an important experimental challenge.
Spontaneous wavefunction collapse models \cite{bassi2003dynamical,bassi2013models,carlesso2022present}  offer a possible answer to it.
They introduce a consistent and minimally invasive modification to the Schr\"odinger equation in order to account for the loss of macroscopic quantum superpositions, by adding non-linear and stochastic terms. Their effect is negligible on microscopic systems, thus preserving their quantum properties, while it becomes stronger for macroscopic systems, causing a progressive breakdown of the quantum superposition principle.
The most studied model is the Continuous Spontaneous Localisation (CSL) model \cite{ghirardi1990markov,pearle1976reduction,pearle1989combining}. This is characterised by two phenomenological constants: the collapse rate $\lambda$, and the spatial resolution of the collapse $\rC$.
There are two main theoretical predictions for these constants, the first one $\rC=10^{-7}\,\mathrm{m}$ and $\lambda=10^{-16}\,\mathrm{s^{-1}}$ proposed by Ghirardi, Rimini and Weber \cite{ghirardi1986unified} and the second one $\lambda\sim10^{-8\pm 2}\;\mathrm{s^{-1}} $ for $\rC=10^{-7}$\,m and $\lambda=10^{-6\pm 2}\;\mathrm{s^{-1}} $ for $\rC=10^{-6}$\,m proposed by Adler \cite{adler2007lower}.
Since this is a phenomenological model, the values of these constants need to be validated through experiments
 \cite{carlesso2022present}. The stronger bounds on the CSL parameters come from non-interferometric class of experiments \cite{bahrami2014proposal,nimmrichter2014optomechanical,diosi2015testing, carlesso2022present}. Such tests aim at detecting the Brownian-like motion, which is induced by the collapse on all systems \cite{donadi2023collapse}. 
 
 \sandro{As shown in~\cite{donadi2023collapse}, 
 such a Brownian-like motion
is a general feature appearing in all models imposing a collapse in space. Thinking in terms of discrete collapses in time, they never occur precisely around the mean value of the  position of the (center of mass of the) system; this means that the mean position (slightly) changes over time, and in the continuous case this changes amount to a diffusion process. When rotating systems are considered, this results in a diffusion in the torque.}
 
In this work, inspired by the experiment in Ref.~\cite{experiment}, we study the CSL effects on the rotational dynamics of a macroscopic optomechanical system. 
The setup in Ref.~\cite{experiment} contains some features that are known to improve the CSL effect: it consists of a macroscopic system, therefore it exhibits the amplification mechanism built in collapse models \cite{carlesso2016experimental}, and it has a periodic mass distribution, which magnifies the collapse in specific regions of the parameter space \cite{Carlesso_2018,diosi2021two}.
\ario{Finally, we analyse the rotational dynamics of the system as it should ensure the experimental advantage of having a low noise environment. Indeed, the rotational degrees of freedom have a much weaker coupling to seismic and acoustic noise than the translational ones \cite{bose2023massive}.}

We find that the experiment in Ref.~\cite{experiment} provides a bound on CSL parameters ($\lambda\simeq10^{-9}\,\mathrm{s^{-1}}$ at $\rC=10^{-4}\,\mathrm{m}$) which is just about one order of magnitude weaker than that derived from the more sophisticated experiment LIGO \cite{carlesso2016experimental}.
Moreover, by suitably modifying the parameters of the experiment, one could be able to push the bounds down to $\lambda\simeq3\times10^{-14}\, \mathrm{s^{-1}}$ at $\rC\simeq10^{-7}\, \mathrm{m}$. This is a bound comparable to that obtained from the collapse-induced radiation emission compared against the data measured in the \textsc{Majorana Demonstrator} experiment on double $\beta$ decay \cite{arnquist2022search} and it becomes the strongest bound at $\rC=10^{-7}\,$m in the case of the (more realistic) non-Markovian (colored) version of the CSL model \cite{adler2007collapse}.

\section{Collapse dynamics and rotations}
\label{sectionII}
The dynamics of the CSL model is given by a master equation \cite{bassi2003dynamical} for the statistical operator of the Lindblad type: $\D \hat{\varrho}(t) / \D t=-\frac{i}{\hbar}[\hat{H}, \hat{\varrho}(t)]+\mathcal{L}[\hat{\varrho}(t)]$, where $\hat{H}$ describes the standard evolution of the system and
\begin{equation}
    \mathcal{L}[\hat{\varrho}(t)]=-\frac{\lambda}{2 \rC^3 \pi^{3 / 2} m_0^2} \int \D \mathbf{z}[\hat{M}(\mathbf{z}),[\hat{M}(\mathbf{z}), \hat{\varrho}(t)]],
    \label{Lindblad}
\end{equation}
accounts for the CSL effects. Here, $m_0$ is a reference mass chosen equal to the mass of a nucleon, $\hat{M}(\mathbf{z})= \sum_n m_n \exp \left(-\frac{\left(\mathbf{z}-\hat{\mathbf{q}}_n\right)^2}{2 \rC^2}\right)$ is a Gaussianly smeared mass density operator, the sum running over the particles of mass $m_n$ of the system. Since the mass of the electron is much smaller than that of nucleons, we can safely consider only the latter, thus setting $m_n=m_0$.

We consider a system whose motion is purely rotational.
In the approximation of small rotations of the system under the action of the CSL noise, $\mathcal{L}[\hat{\varrho}(t)]$ can be expanded around the equilibrium angle \cite{schrinski2017collapse}.
In this case, Eq.~\eqref{Lindblad} reduces to:
\begin{equation}
    \mathcal{L}[\hat{\varrho}(t)]=-\frac{\eta}{2}\left[\hat{\theta},\left[\hat{\theta}, \hat{\varrho}(t)\right]\right],
    \label{Lindbladapproximation}
\end{equation}
where $\hat{\theta}$ is the angular operator describing rotation around a fixed axis and $\eta$ is a function of the mass density of the system.

Following the idea developed in Ref.~\cite{carlesso2018multilayer},  we explore how to enhance the CSL effect in this purely rotational case by optimizing the shape and the mass density distribution of an hypothetical test mass. \ario{Once the geometry of the system is suitably chosen, the rotational degrees of freedom are more advantageous than the translational ones since the first are subject to less environmental noises.}
\ario{In this work} the choice of the shape is inspired by the disk used as a torsion balance reported in the experiment in Ref.~\cite{experiment}, which is depicted in Fig.~\ref{fig:geometry}(a). 
The equations of motion of the pendulum we are investigating read \cite{carlesso2016experimental} 
\begin{equation}\label{eq.langevin}
	\begin{aligned}
		&\dfrac{\D\hat{\theta}}{\D t}=\frac{\hat{L}}{I},&\dfrac{\D\hat{L}}{\D t}= -I\omega_0^2\hat{\theta}-\gamma \hat{L} +\hat{\tau}_\text{th} +\hat{\tau}_\text{\tiny CSL},
	\end{aligned}
\end{equation}
where $\omega_0$, $\gamma$, and $I$ are, respectively, the resonance frequency of the torsion balance, the damping of the resonator, and the moment of inertia of the system;  $\tau_\text{th}$  and $\tau_\text{\tiny CSL}$ are the thermal and the CSL stochastic torques.
A complete treatment of the problem should consider extra noise terms due to the measurement. However, we take a conservative approach, and assume that all non-thermal noises are caused by CSL. Accounting for other noises can only improve the bounds on the CSL parameters.

Once the correlation functions of the two torques are evaluated, one can derive the thermal and CSL contribution to the Density Noise Spectrum (DNS), whose form is $S_\tau(\omega)= \int_{-\infty}^{\infty} \D s\, e^{-i\omega s} \mathbb{E}[\langle \hat \tau(t)\hat \tau(t+s)\rangle]$, where $\mathbb{E}\left[...\right]$ represents the average over the collapse and on the thermal noise, while $\langle...\rangle$ the standard quantum average.
A common experimental design involves monitoring the position or the rotation of the system and then determining the force exerted on it, expressing it in terms of the DNS.
In the case of a mass with cylindrical symmetry rotating around its axis, the CSL contribution to the torque DNS has the following expression (see Appendix \ref{appA}):
\begin{equation}
    S_\text{\tiny CSL}(\omega)=\frac{\lambda\hbar^2}{4m_0^2\rC^4} P \times Y ,
    \label{DNS}
\end{equation}
with
\begin{equation}
    \begin{aligned}
       &Y=\int_{-h/2}^{h/2} \D y\int_{-h/2}^{h/2} \D y' e^{-\frac{(y-y')^2}{4 \rC^2}},\\
       &P=\int_0^{\infty} \D r_\perp  \int_0^{\infty} \D r'_\perp  r^2_\perp r'^2_\perp e^{-\frac{r_\perp^2+r_\perp'^2}{4 \rC^2}} A(r_\perp,r_\perp'),
    \end{aligned}
\end{equation}
where the integrals are expressed in term of the cylindrical coordinates $\left(r_\perp,\theta,y\right)$, with $r_\perp$ and $\theta$ determining the points of the plane represented in Fig.~\ref{fig:geometry} and $y$ the perpendicular direction.
Moreover, we assume a mass distribution of the form $\varrho\left(r_\perp,\theta,y\right)=H(h/2-y)H(y-h/2)\varrho_{P}(r_\perp,\theta)$ expressed in terms of the Heaviside function $H$, with $h$ being the thickness of the cylinder.
Finally, we define
\begin{equation}
    \begin{aligned}
    &A(r_\perp,r_\perp')=\int_0^{2\pi} \D \theta \int_0^{2\pi} \D \theta' \varrho_{P}(r_{\perp},\theta)\varrho_{P}(r'_{\perp},\theta')  \times\\
    &\times \left(2 \rC^2 \cos (\theta -\theta')-r_\perp r_\perp' \sin ^2(\theta -\theta')\right) e^{-\frac{2 r_\perp r_\perp' \cos (\theta -\theta')}{4 \rC^2}}\;,
    \label{St}
    \end{aligned}
\end{equation}
which explicitly accounts for the angular and radial mass distribution. To derive a bound, we compare the contribution to the spectral density due to the CSL $S_\text{\tiny CSL}(\omega)$  with that due to thermal fluctuation{, which reads $S_\text{th}(\omega) = 4\kb T\gamma I$}.
The bound is found by imposing $S_\text{\tiny CSL}(\omega)\leq S_\text{th}(\omega)$, this is a conservative approach in which we assume that the CSL contribution is responsible at most for the entire thermal contribution to the DNS.
If we modify the mass density of the system without altering the moment of inertia $I$, then $S_\text{th}$ remains constant.
For this reason most of the analysis here is performed by keeping $I$ fixed.

\section{A simplified model}
\begin{figure}[t!]
    \centering
        \includegraphics[width=\linewidth]{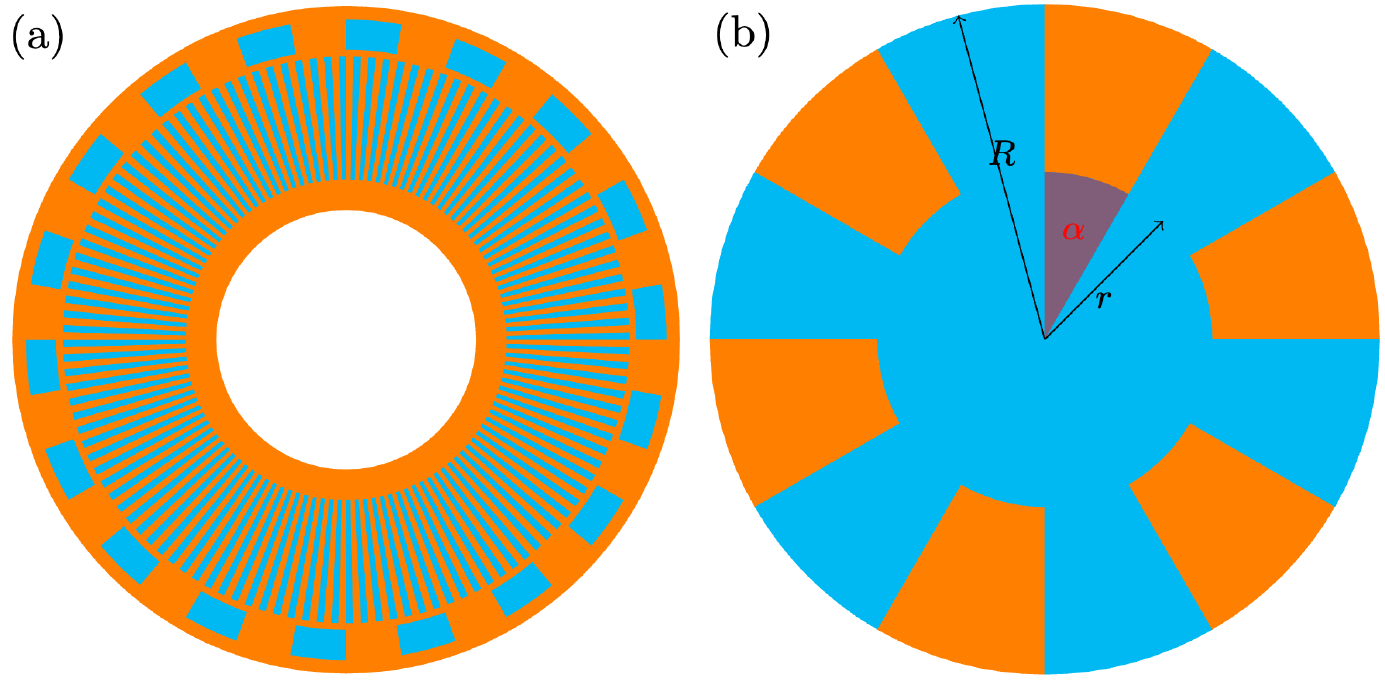}
    \caption{Top view of the mass density configurations. The orange region has density $\varrho+\Delta\varrho$ and the cyan one has density $\varrho$. Panel (a) shows the actual configuration used in \cite{experiment}. Panel (b) represents a simplified configuration; the angle subtended by the orange sectors is denoted by $\alpha$, while $r$ and $R$ represent the inner and the outer radii of the annulus with periodic mass density. 
    }
    \label{fig:geometry}
\end{figure}

The following analysis aims at enhancing the CSL effect by introducing a periodic mass density in the angular variable as depicted in Fig.~\ref{fig:geometry}(b). Here, we study a simplified model, where the system is composed by a single annulus with a periodic mass density. This is a case with fewer parameters with respect to that in Fig.~\ref{fig:geometry}(a), which refers to the experiment in Ref.~\cite{experiment}.
To evaluate the torque DNS, we 
introduce the following mass density function for this configuration:
\begin{equation}
		\begin{aligned}    
   &\varrho_{P}(r_{\perp},\theta)=H(R-r_{\perp})\bigg[\varrho H(r_{\perp})+\Delta\varrho H(r_{\perp}-r)\\
            &\times\sum_{j=0}^{n-1} H\left(\theta-\frac{2j\pi}{n}\right)H\left(\frac{2j\pi}{n}+\alpha-\theta\right)\bigg] ,
		\end{aligned}
\end{equation}
where $\varrho$ is the mass density of the lighter material (shown in cyan in Fig.~\ref{fig:geometry}), $\Delta\varrho>0$ is the difference of mass densities between the two materials, $2n$ is the number of sectors in which the annulus is divided, $\alpha\in\left[0,2\pi/n\right]$ is the angle subtended by a orange sector, $r$ and $R$ are respectively the inner and outer radii. 
We show in Appendix \ref{appA} that the first term in parenthesis
, corresponding to the homogeneous cylinder at the centre of the annulus, does not contribute to the CSL effect. Conversely, the second one does. Thus, according to Eq.~\eqref{St}, the effect scales with the square of the density difference between the materials $\Delta \rho$.
Moreover, in the configuration just described, Eq.~\eqref{St} can be evaluated analitically leading to the following expression
\begin{figure*}[th]
\centering
    \includegraphics[width=1\linewidth]{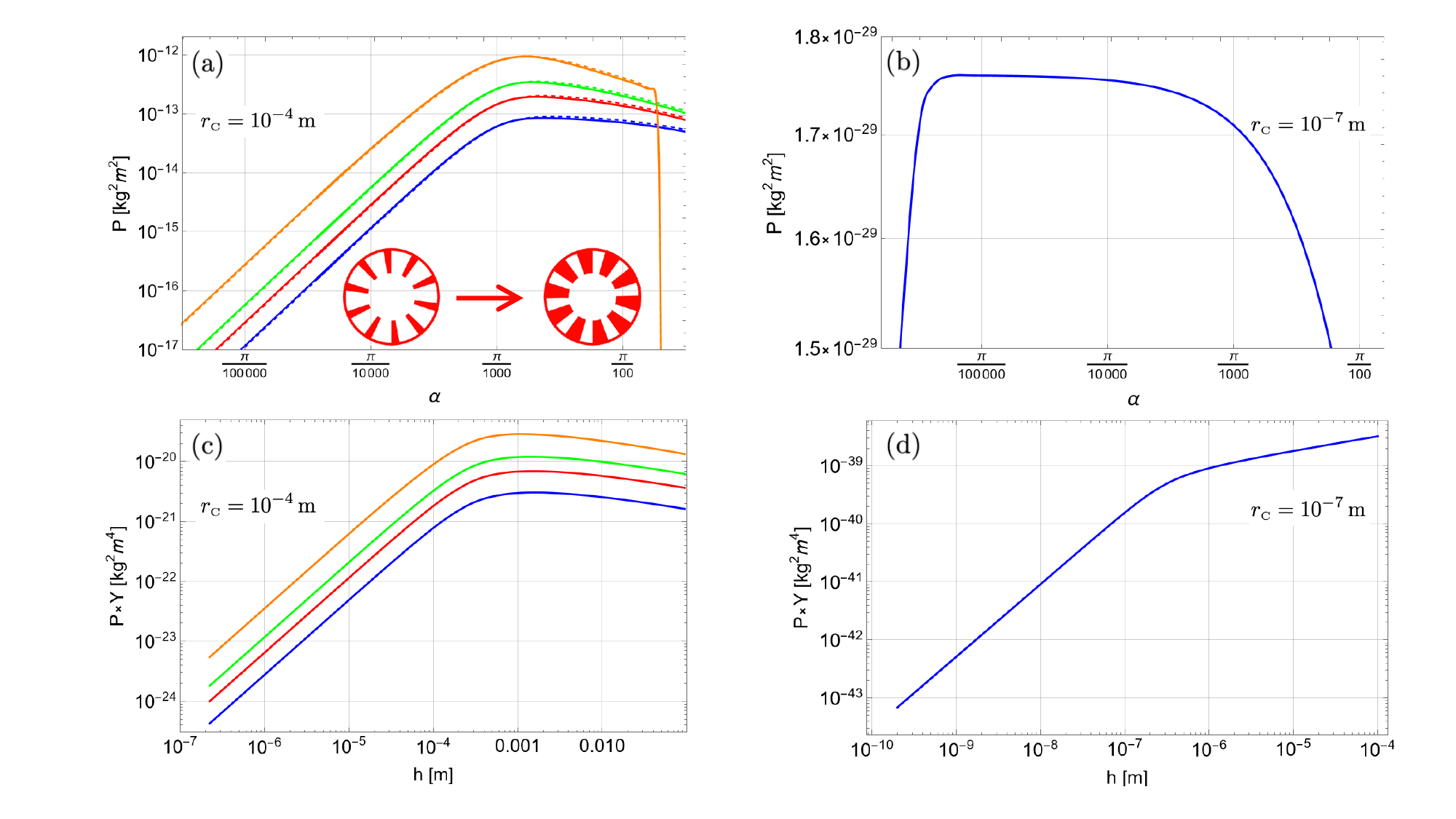}
    \caption{Upper Panels: comparison of the values of $P$ for different mass density functions of the simplified model for (a) $\rC=10^{-4}\,\mathrm{m}$ and  (b) $\rC=10^{-7}\,\mathrm{m}$ . The values of $n$ taken into consideration are $n=4$ (blue lines), $n=10$ (red lines), $n=20$ (green lines), $n=100$ (orange lines). The orange line in panel (a) drops at $\alpha=2\pi/100$, which corresponds to the homogeneous limit. The solid and dashed lines represent the cases in which $\epsilon=2$ and $\epsilon=20$, respectively. The moment of inertia is kept fixed at $I=9\times10^{-6}\,\mathrm{kg}\,\mathrm{m^2}$ by changing the value of 
    $r$. The thickness of the cylinder is kept fixed at $h=10^{-4}\,\mathrm{m}$.
    Lower panels: comparison of the values of $P\times Y$ for different mass density functions of the simplified model for (c) $\rC=10^{-4}\,\mathrm{m}$ and (d) $\rC=10^{-7}\,\mathrm{m}$. 
    The parameter $h$ is varying  while $\epsilon=2$ and $I=9\times 10^{-6}\,\mathrm{kg}\,\mathrm{m^2}$ are kept fixed. In panels (b) and (d) we only analysed the case with $n=4$ since for $\rC=10^{-7}\mathrm{m}$ the effect of the model is much smaller than the case of $\rC=10^{-4}$\,m, and it is not worth a more detailed analysis.
    \label{fig:Srt}}
\end{figure*}
\begin{equation}
    \begin{aligned}
A(r_\perp,r_\perp')=&\frac{32\rC^4 \Delta\varrho^2}{r_\perp r_\perp'}\prod_{r_x=r'_{\perp},r_{\perp}}H(r_x-r)H(R-r_x) \\
&\times 
\sum_{j=0}^{\infty}I_{j n}\left(\frac{r_\perp r'_\perp}{2\rC^2}\right)   n^2 \sin ^2\left(\frac{\alpha  j n}{2}\right),
    \end{aligned}
    \label{St1}
\end{equation}
where $I_\sigma(x)$ are the modified Bessel function of the first kind of order $\sigma=j\times n$.
We note that $A$ is zero for $\alpha=0$ and $\alpha=2\pi/n$, since these value correspond to a homogeneous mass density configuration.

We can start our numerical analysis of $Y$ and $P$ noting that, once the value of the moment of inertia $I$ and the material densities are fixed, $P$ depends on: the angle $\alpha$, the number $n$ of heavier (orange) sectors, the inner ($r$) and outer ($R$) radii and the height ($h$).
\ario{This choice follows that performed in Ref.~\cite{carlesso2018multilayer}, where the translational moment of inertial, namely the mass, was fixed. Such a choice provides a fair comparison between the proposed configurations, since the value of $S_\text{th}$ does not change.}
We consider the values $\rC=10^{-4}\,\text{m}$ and $\rC=10^{-7}\,\text{m}$ for reference, and compute $P$ for $n=4,10,20$ and  $\epsilon={R}/{r}=2$ and $20$ by 
varying $\alpha$ in the interval $[0,2\pi/n]$.
To keep the value of $I$ fixed, we change the value of $r$ as a function of the different values assumed by $n$, $\alpha$ and $\epsilon$, while keeping $h$ fixed.
Under this assumption, the value of $Y$ is constant.

Figure \ref{fig:Srt} 
shows the dependence of $P$ from $\alpha$ for $\rC\simeq10^{-4}\,\mathrm{m}$ and $\rC\simeq10^{-7}\,\mathrm{m}$. The optimal value of $\alpha$ does not depend strongly on the value of $n$, while it does on the value of $\rC$.
This behaviour is the same that has been noticed in Ref.~\cite{carlesso2018multilayer}. Indeed, in this case the maximum of the CSL effect occurs when $\rC$ and  the arc length subtended by the $\alpha$ sector are similar.
Panel (a) shows that there is an enhancement of the CSL effect on $P$ for $\rC=10^{-4}\,$m when increasing the number $n$ of sectors. The dependence on $\epsilon$ seems less impactfull:  the dashed line ($\epsilon=20$) and the solid line ($\epsilon=2$) almost completely overlap for every $n$. Finally, it is important to note that the orange curve, corresponding to the annulus with $n=100$ orange sectors, goes to zero for $\alpha \rightarrow 2\pi/100$. This is expected, since it corresponds to an homogeneous mass configuration.
Panel (b) shows the behaviour of $P$ for $\rC=10^{-7}\,\mathrm{m}$: the constraints imposed on the geometry (the choice of the value of $r$) produce a weaker CSL effect in comparison with that shown in panel (a).

In the following numerical analysis, we fix the values of $\rC$, $\epsilon$,  $I$ and therefore $\alpha$ to their optimal values, based on the choice of $\rC$.
These are shown in Tab.~\ref{tab:my_label}.
We then evaluate $P\times Y$ by letting $h$ and $n$ vary, and at the same time we change $r$ to maintain $I$ constant. This analysis gives us the optimal value of $h$ from panel (c) and (d) of Fig.~\ref{fig:Srt} in correspondence of the two values of $\rC$ here considered. 
\renewcommand{\arraystretch}{1.2}
\begin{table}[b!]
    \centering
    \begin{tabular}{c|c|c|c|c|c}
        $\rC\,[\mathrm{m}]$ & $\epsilon$ & $I\,[\mathrm{kg}\,\mathrm{m^2}]$ & $\alpha$ & $h\,[\mathrm{m}]$&n\\
        \hline
         $10^{-4}$ & $2$ & $9\times10^{-6}$ &$5\times 10^{-3}$&$10^{-3}$& 100  \\
         \hline
         $10^{-7}$ & $2$ &$9\times10^{-6}$ & $3\times10^{-5}$ & $ 6\times10^{-3}$&4\\
    \end{tabular}
    \caption{Optimised values of the parameters obtained from the numerical analysis for the simplified model.}
    \label{tab:my_label}
\end{table}

\begin{table*}[t]
\centering
 \ario{\begin{tabular}{c|c|c|c|c}
     $h\,[\mathrm{m}]$ & $S_\text{th}[\mathrm{N^2}\,\mathrm{m^2}\,\mathrm{s}]$& $\rho\,[\mathrm{kg/m^3}]$ & $\Delta\rho\,[\mathrm{kg/m^3}]$ &$r\,[\mathrm{m}]$  \\
    \hline
     $5.4\times 10^{-5}$ & $1\times10^{-30}$ & $1.2\times10^{3}$ & $19.3\times10^{3}$ &$1.05\times10^{-2}$  \\
     \hline
	 $r_{120}\,[\mathrm{m}]$ & $R_{120}\,[\mathrm{m}]$ & $r_{18}\,[\mathrm{m}]$ & $R_{18}\,[\mathrm{m}]$ & $R\,[\mathrm{m}]$  \\
    \hline
	  $1.30\times10^{-2}$ & $2.30\times10^{-2}$ & $2.35\times10^{-2}$ & $2.60\times10^{-2}$ & $2.70\times10^{-2}$  \\ 
	\hline
	\end{tabular}
\caption{Parameters of the setup in \cite{experiment} (see also Fig.~\ref{fig:geometry}): the height $h$, thermal contribution to the noise $S_\text{th}$, mass density $\rho$, mass density difference $\Delta\rho$, and the radii (from the most inner to the most outer): internal radius $r$, for the annulus with 120 sectors internal radius $r_{120}$ and  external $R_{120}$, for the annulus with 18 sectors internal radius $r_{18}$ and external $R_{18}$, external radius $R$.}}
\label{table}
\end{table*}

In summary, from Fig.~\ref{fig:Srt} it is possible to identify the optimal values of $\alpha$ and $h$ in order to enhance the bound at $\rC=10^{-4}\,\mathrm{m}$ and $\rC=10^{-7}\,\mathrm{m}$. Fixing these parameters to their optimal value [cf.~Tab.~\ref{tab:my_label}] for $\rC=10^{-4}\,\mathrm{m}$, 
we derive the corresponding bound, which is reported in Fig.~\ref{fig:BOUND} (black line).

\section{Comparison with experimental data}
The experiment \cite{experiment} that inspired this analysis was designed to test gravity over short distances to find possible violations of the gravitational inverse square law.
In particular, the experiment was used to constrain a possible additional Yukawa interaction to the Newtonian potential of the form: $
V(r)=V_N(r)[1+a \exp (-r / \ell)]$,
where $V_N(r)$ is the Newtonian potential, and $a$ and $\ell$ are the free parameters to be tested. The mass used in the experiment is represented in Fig.~\ref{fig:geometry}(a).

\begin{figure}[th]
    \centering
    \includegraphics[width=0.9\linewidth]{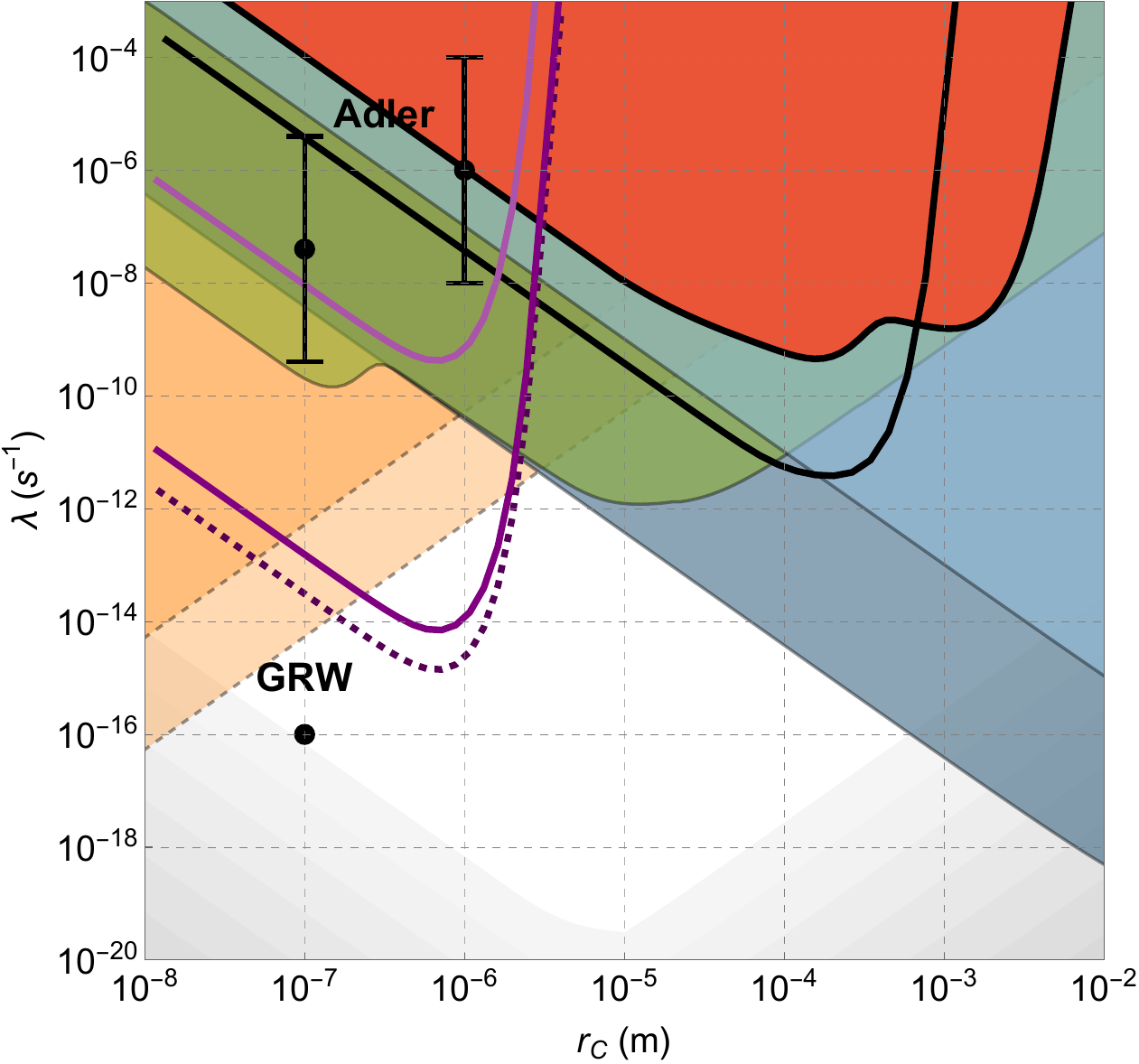}
    \caption{Exclusion plot for the CSL parameters $\lambda$ and $\rC$ from rotational tests, compared with the existing experimental bounds. The red area represent the region excluded by the experiment in Ref.~\cite{experiment}. The black line is the hypothetical upper bound derived by optimising the geometry of the simplified model for $\rC\simeq10^{-4}\,\mathrm{m}$ and \ario{$T=300\,\mathrm{K}$}.
    The  purple lines refers to the hypothetical bounds derived by optimizing the geometry of the simplified model for $\rC\simeq10^{-6}\,\mathrm{m}$ at different temperatures $T$, respectively  \ario{$T=300\,\mathrm{K}$ (continuous light purple line)}, \ario{$T=50\,\mathrm{mK}$ (continuous dark purple line) and $T=1\,\mathrm{mK}$ (dashed dark purple line)}. 
    These upper bound are compared with the coloured areas corresponding to regions already excluded experimentally. 
    The green regions refer to
    cantilever-based experiments with multilayer structures \cite{Vinante_2020}. The blue areas to gravitational wave detectors \cite{carlesso2016experimental,carlesso2018non,helou2017lisa}: LIGO (light blue) and LISA Pathfinder (dark blue). The orange areas delimited by the dashed lines are from spontaneous $\mathrm{X}$-ray emission tests: the darker one is derived in Ref.~\cite{donadi2021novel} and the lighter one is derived with data from the \textsc{Majorana Demonstrator} \cite{arnquist2022search}.
    \sandro{The region excluded by theoretical requirements is represented in grey, and it is obtained by requiring that macroscopic superposition should not persist in time (see appendix \ref{appD}).} The white area is yet to be explored.}
    \label{fig:BOUND}
\end{figure}

The disk considered in the experiment consists of two concentric annuli; we then have extended the analysis presented in the previous section to more then one annulus. In this way we will show that it is possible to enhance the CSL effect for different values of $\rC$ simultaneously. In the case of two concentric annuli with different angular periodicity (internal annulus with $n$ orange sectors, external annulus with $m$ orange sectors [cf.~Fig.~\ref{fig:geometry}(a)]) Eq.~\eqref{St} takes the following form:
\begin{widetext}
\begin{equation}\label{St2}
		\begin{aligned}
			&A(r_\perp,r_\perp',m,n)=\frac{32 \rC^4}{r_\perp r'_\perp}\Delta \varrho^2 \sum_{k=1}^{\infty} I_k\left(\frac{r_\perp r'_\perp}{2\rC^2}\right)\bigg[\sum_{\nu=n,m} \nu^2\sum_{h=0}^{\infty}\delta_{k,(2h+1)\nu} \prod_{r_x=r'_{\perp},r_{\perp}}H(r_x-r_{\nu})H(R_{\nu}-r_x)+\\ 
            &+ mnH(r'_\perp-r_{n})H(R_{n}-r'_\perp)H(r_{\perp}-r_{m})H(R_{m}-r_{\perp})\bigg(\sum_{h=0}^{\infty}\sum_{h'=0}^{\infty}\delta_{k,(2h+1)m}\delta_{k,(2h'+1)n}\bigg)\bigg],
		\end{aligned}
	\end{equation}
 \end{widetext}
where $r_\nu$ and $R_\nu$ are respectively the inner and outer radii of the $\nu$-th annulus, with $\nu=n$ and $\nu=m$ indicating the internal and external annulus respectively.
We recall that the terms  representing an homogeneous mass density do not contribute to the effect.
In the second line of Eq.~\eqref{St2} a mixed term is present, this is where both the inner and outer annuli parameters appear. It vanishes if $n$ and $m$ satisfy the condition $(2k+1)m\not=(2k'+1)n,\ \forall k\in \mathbb{N},\forall k'\in \mathbb{N}$.
In our case, we have $n=120$ and $m=18$ that satisfy this condition, thus only the first line of Eq.~\eqref{St2} contributes.
Now we take the experimental results reported in Ref.~\cite{experiment} to set an upper bound on the  parameters of the CSL model as discussed in Sec.~\ref{sectionII}. In doing this, we consider the frequency region between $2\times 10^{-3}\,$Hz and $10^{-1}\,$Hz (the resonant frequency is $\omega_0=1.8\times 10^{-2}\,$Hz) of the experimental spectrum in which the main noise is the thermal one. The parameters characterizing the test mass are summarised in Tab.~\ref{table}.
The corresponding bound  is shown in Fig.~\ref{fig:BOUND} with the red area: it has two local minima reflecting the two different periodicities.

\section{Discussion and future perspective}
\label{Discussion and future perspective}
In the relevant range of values of $\rC$, the bound derived from the experiment in Ref.~\cite{experiment} 
is comparable to that excluded by the much more sophisticated experiment LIGO \cite{carlesso2016experimental}.
The considered experiment was not designed to test the CSL model, therefore it is possible to optimize the geometry of the system to improve even further the bound in regions of the parameters plot yet to be explored.
For example, in the simplified model, it is possible to derive a bound with its minimum at $\rC=10^{-6}\,\mathrm{m}$  by choosing the parameters as follows: $r=10^{-5}\,\mathrm{m}$, $R=2\times10^{-4}\,\mathrm{m}$, $n=300$, $\alpha=\pi/n$, $h=10^{-3}\,\mathrm{m}$.
If we assume \ario{$T=300\,\mathrm{K}$}, we obtain the bound represented with a light purple line.
However, by taking \ario{$T\simeq50\,\mathrm{mK}$}  we obtain a stronger bound (dark purple line), which allows to explore a new region at $\rC=10^{-6}\,\mathrm{m}$. \ario{The corresponding thermal contribution to the noise has been obtained by rescaling the experimental thermal noise with respect to the ratio of the moment of inertia and temperature, namely $S_\text{th,new}=S_\text{th,exp}(I_\text{new}T_\text{new})/(I_\text{exp}T_\text{exp})$. 
This bound }comes with the assumptions that it is possible to realize a test mass with the above parameters, and that thermalises at \ario{$50\,\mathrm{mK}$}. Experiments around \ario{the temperature of $T=50\,\mathrm{mK}$} have already been carried out \cite{Vinante_2020}. \matteo{An even stronger bound (dashed purple line) can be obtained with  $T\simeq1\,\mathrm{mK}$, which is a temperature that it is reasonable to expect can be achieved in near-future experiments.} 

To conclude, we summarise the main properties of the proposed technique: a geometry with concentric annuli is capable of simultaneously probing multiple regions of the parameter space. The effect of the model is maximum approximately when the arc length subtended by the sectors is comparable to $\rC$ (this is verifiable analytically for a simple case discussed in Appendix \ref{appB}). Conversely, there is no advantage into applying this technique to test $\rC$ bigger than the system's dimension. Indeed, the effect fades rapidly as $\rC$ increases. Our analysis shows that in principle this technique can offer competitive bounds in the same region ($10^{-7}\;\mathrm{m}<\rC<10^{-6}\;\mathrm{m}$) as that touched by X-ray detection experiments (orange areas in Fig.~\ref{fig:BOUND}). However, the latter experiments, in contrast to mechanical oscillators, target the high frequency region of the CSL noise spectrum and as such are much more sensible to changes in noise, for example based on the introduction of a cutoff \cite{adler2013spontaneous,donadi2014spontaneous,Carlesso_2018}. In such a case, the bounds highlighted in orange loose strength and the purple bound presented here becomes the dominant one. 

\section*{Acknowledgments}
The authors acknowledge the UK
EPSRC (Grant No.~EP/T028106/1), {the EU EIC Pathfinder project QuCoM (101046973)}, the PNRR PE National Quantum Science and Technology Institute (PE0000023), the  Marie Sklodowska Curie Action through the UK Horizon Europe guarantee administered by UKRI, the University of Trieste and INFN.

\vspace*{\fill}

   \onecolumngrid
   \newpage
\appendix
\section{CSL Torque from the Master Equation}
\label{appA}
We derive the CSL torque $\tau_{\text{\tiny CSL}}$ starting from the master equation \eqref{Lindblad}.
This dynamics can be reproduced by a standard Schrödinger equation with an additional stochastic potential of the form \cite{fu1997spontaneous,carlesso2016experimental}:
\begin{equation}
    \hat{V}_{\text{\tiny CSL}}(t)=-\frac{\hbar \sqrt{\lambda}}{\pi^{3 / 4} \rC^{3 / 2} m_0^2} \int \D \boldsymbol{z} \hat{M}(\boldsymbol{z}) w(\boldsymbol{z}, t),
\end{equation}
where $w(\boldsymbol{z}, t)$ is a collection of white noises (one for each point of space $\boldsymbol{z}$) with $\mathbb{E}[w(\boldsymbol{z}, t)]=0$ and $\mathbb{E}[w(\boldsymbol{z}, t) w(\boldsymbol{y}, s)]=\delta(t-s) \delta^{(3)}(\boldsymbol{z}-\boldsymbol{y})$. Such a stochastic potential acts on the $n$-th particles of the system as a stochastic force:
\begin{equation}\label{Fn}
    \boldsymbol{\hat F}_n=\frac{i}{\hbar}[\hat V_{\text{\tiny CSL}},\boldsymbol{\hat p}_n].
\end{equation}
Then the position operator can be written as $\boldsymbol{\hat q}_n= \boldsymbol{q}^{(0)}_n+\Delta \boldsymbol{\hat q}_n$, where $\boldsymbol{q}^{(0)}_n$ is the classical equilibrium position of the $\alpha$-th nucleon and $\Delta \boldsymbol{\hat q}_n$ quantifies the quantum displacement of the $n$-th nucleon with respect to its classical equilibrium position. Now assuming that we are dealing with a rigid body and that the quantum fluctuations are small with respect to $\rC$, we can Taylor expand the mass density:\ario{ 
\begin{equation}
    \hat M(\boldsymbol{z})=M_0(\boldsymbol{z})+\rC^{-2} \int \D \boldsymbol{x} \,\varrho(\boldsymbol{x}) \exp\left(-(\boldsymbol{z}-\boldsymbol{x})^2/\left(2 \rC^2\right)\right)(\boldsymbol{z}-\boldsymbol{x}) \cdot \Delta\hat{\boldsymbol{q}},
\end{equation}}
\ario{where $\Delta \hat{q}$ represents the fluctuation of the position of the center of mass} and $M_0(z)$ is a classical function whose form is not important. Thus, Eq.~\eqref{Fn} becomes:
\begin{equation}
   \boldsymbol{ F}_n(t)=\frac{\hbar \sqrt{\lambda}}{\pi^{3 / 4} m_0} \int \frac{\D \boldsymbol{z} }{\rC^{7 / 2}} e^{-\frac{(\boldsymbol{z}-\boldsymbol{ q}^{(0)}_n)^2}{2 \rC^2}}(\boldsymbol{z}-\boldsymbol{ q}^{(0)}_n) w(\boldsymbol{z}, t).
\end{equation}
Since the geometry of the system studied is cylindrical, as depicted in Fig.~\ref{fig:geometry}, it is easier to handle the problem using cylindrical coordinates, defined as $y$, $r_\perp=\sqrt{x^2+z^2}$ and $\theta=\arctan(x/z)$.
By using the tangent component of the force $F_{\theta}(\boldsymbol{x},t)=\boldsymbol{F}(\boldsymbol{x},t)\cdot \boldsymbol{e}_{\theta}$, we can evaluate the torque along $y$ acting on the whole system:
\begin{equation} 	
\tau_\text{\tiny CSL}\left(t\right)=\int\,\D r_\perp \D\theta \D y\; r_\perp^2\,F_\theta(\boldsymbol{x},t).
\end{equation}
Starting from the correlation function for $F_{\theta}(\boldsymbol{x},t)$
\begin{equation}
	\begin{aligned}
	& \mathbb{E}[F_\theta(\boldsymbol{x},t)\,F_\theta(\boldsymbol{x}',s)]=\frac{\lambda\hbar^2}{4m_0^2\rC^4}\left(2 \rC^2 \cos (\theta -\theta')-r_\perp r_\perp' \sin ^2(\theta -\theta')\right)\varrho(\boldsymbol{x})\varrho(\boldsymbol{x}')\times\\
	&\times  \exp\left(-\frac{r_\perp^2-2 r_\perp r_\perp' \cos (\theta -\theta')+r_\perp'^2+(y-y')^2}{4 \rC^2}\right)\delta(t-s),
 \label{correlationFtheta}
	\end{aligned}
\end{equation}
we can evaluate the correlation function for $\tau_\text{\tiny CSL}$ as:
\begin{equation}
    \begin{aligned}
    &\mathbb{E}\left[\tau_\text{\tiny CSL}(t)\tau_\text{\tiny CSL}(s)\right]=\int\,\D r_\perp \D \theta 
\D y\;\int\,\D r'_\perp \D \theta'\D y'\;r^2_\perp r'^2_\perp \;\mathbb{E}[\,F_\theta(\boldsymbol{x},t)\,F_\theta(\boldsymbol{x}',s)].
\label{0}
    \end{aligned}
\end{equation}
Finally, one obtains the DNS via:
\begin{equation}
    S_\text{\tiny CSL}(\omega)=\int_{-\infty}^{\infty}\D s\, e^{-i\omega s} \mathbb E\left[\tau_\text{\tiny CSL}(t)\tau_\text{\tiny CSL}(t+s)\right].
\end{equation}
Now, we analyze the case in which the mass density is independent from $\theta$, i.e.~rotationally homogeneous. We consider only the radial and angular part of the previous integral.
We recall the following known identities:
\begin{equation}
	\int_0^{2\pi}\D \theta \cos (\theta -\theta') \exp\left(\frac{ r_\perp r_\perp' \cos (\theta -\theta')}{2 \rC^2}\right)= 2\pi I_1\left(\frac{ r_\perp r_\perp'}{2\rC^2} \right)
	\label{1}
\end{equation}
and
\begin{equation}
	\int_0^{2\pi}\D \theta \sin ^2(\theta -\theta') \exp\left(\frac{ r_\perp r_\perp' \cos (\theta -\theta')}{2 \rC^2}\right)= 2\pi \frac{I_1\left(\frac{ r_\perp r_\perp'}{2\rC^2} \right)}{\frac{ r_\perp r_\perp'}{2\rC^2}}.
	\label{2}
\end{equation}
where $I_1$ is the modified Bessel function of the first kind. Finally, by replacing Eq.~\eqref{1} and Eq.~\eqref{2} in  Eq.~\eqref{0} one finds that the CSL effect vanishes. This means that CSL has no effect on rotations of a rotationally homogeneous system.

\section{Study of a simple system for understanding the CSL amplification mechanism}
\label{appB}
To better understand the relation between the size of the test mass and the maximization of the CSL effects, we consider the simple example of a half cylinder of radius $R$.
In this case the expression in Eq.~\eqref{St} takes the simple form:
\begin{equation}
   \begin{aligned}
       &A(r_\perp,r_\perp')=32\rC^4 \Delta\varrho^2\sum_{j=0}^{\infty}I_{2j+1}\left(\frac{r_\perp r'_\perp}{2\rC^2}\right)\prod_{r_x=r'_{\perp},r_{\perp}}\frac{H(R-r_x)}{r_x} =16\rC^4 \Delta\varrho^2 \sinh\left(\frac{r_\perp r'_\perp}{2\rC^2}\right)\prod_{r_x=r'_{\perp},r_{\perp}}\frac{H(R-r_x)}{r_x},
   \end{aligned}
\end{equation}
and the corresponding form of $P$ becomes:
\begin{equation}
    P=16\rC^4 \Delta \varrho^2 \int_0^{R} \D r_\perp  \int_0^{R} \D r'_\perp  r_\perp r'_\perp e^{-\frac{r_\perp^2+r_\perp'^2}{4 \rC^2}} \sinh\left(\frac{r_\perp r'_\perp}{2\rC^2}\right).
    \label{semicilindro}
\end{equation}
Then, we equate the CSL contribution to the thermal noise, which depends on the moment of inertia of the half cylinder: $I=\frac{\pi}{4} h \varrho R^4$. 
It follows that the dependence of the corresponding  $\lambda$ can be expressed in terms of  $\lambda\propto (F(R/\rC))^{-1}$ where
\begin{equation}
F\left(\frac{R}{\rC}\right)= \frac{2 - 3 \left(\frac{R}{\rC}\right)^2 + e^{-\left(\frac{R}{\rC}\right)^2} \left(-2 + \left(\frac{R}{\rC}\right)^2\right) + \sqrt{\pi} \left(\frac{R}{\rC}\right)^3 \mathrm{erf}\left(\frac{R}{\rC}\right)}{\left(\frac{R}{\rC}\right)^4}.
\end{equation}

\begin{figure}[h!]
    \centering
    \includegraphics[width=0.5\linewidth]{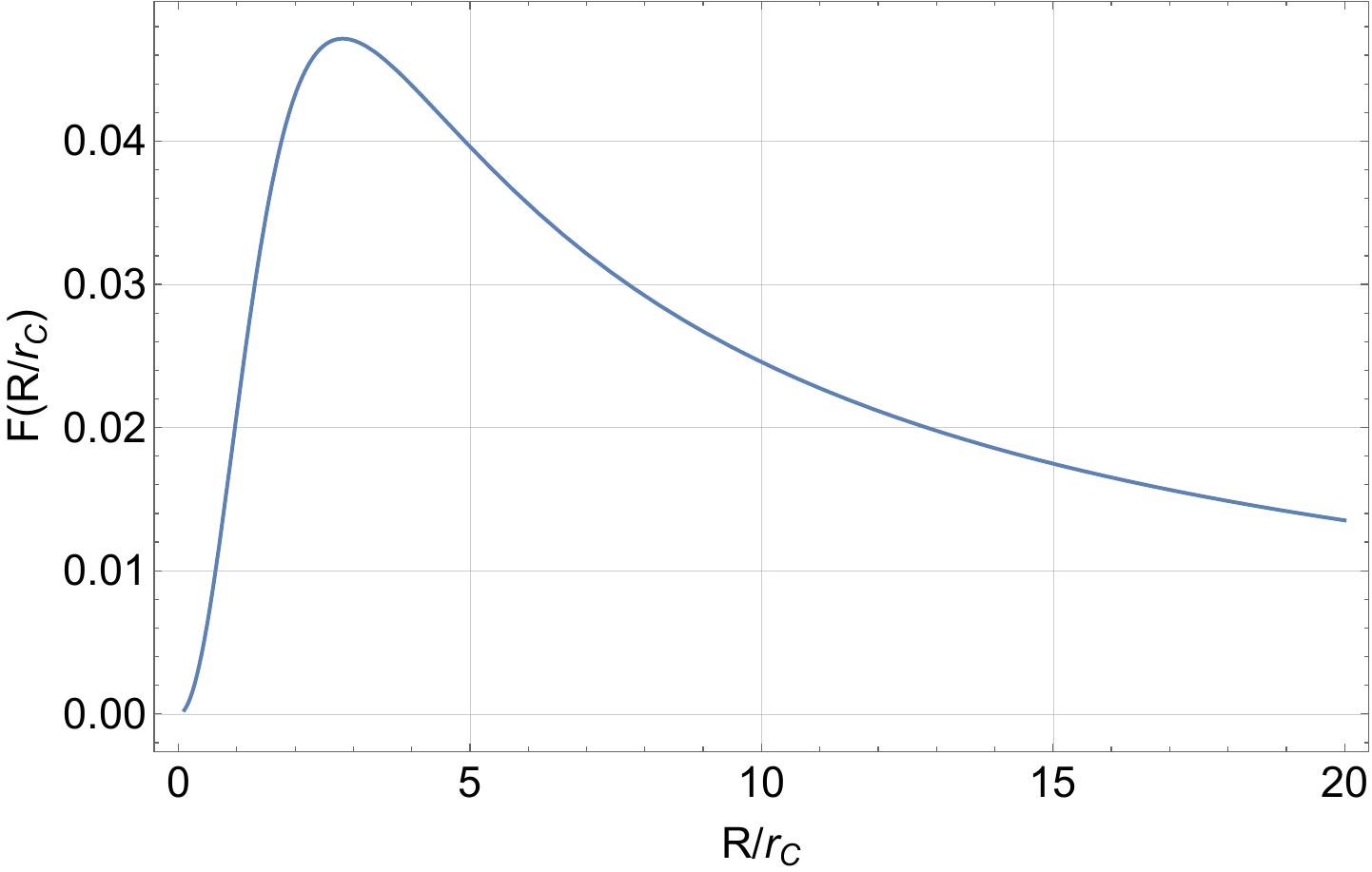}
    \caption{Dependence of $F$ in Eq.~\eqref{semicilindro} on the dimensionless ratio between the radius of the cylinder and the CSL parameter $\rC$}
    \label{fig:Semicilindro}
\end{figure}

From the plot in Fig.~\ref{fig:Semicilindro} we can see that the maximum of the term in parenthesis (which gives the optimized value of $R/\rC$ in order to have a stronger bound for $\lambda$) is for values of $R/\rC \simeq 3 $.

\section{Colored CSL Evaluation}
We can generalise our calculation to the colored version of the CSL model (the quantities relative to this model contain the label C), in which $\mathbb{E}[w_\text{\tiny C}(\boldsymbol{z}, t) w_\text{\tiny C}(\boldsymbol{y}, s)]= f(t-s) \delta^{(3)}(\boldsymbol{z}-\boldsymbol{y})$, where $f(t)$ is a correlation function with colored spectrum. By taking $f(t)=\delta(t)$ one recovers the standard CSL model.
In this case the correlation function for $F^\text{\tiny C}_{\theta}(\boldsymbol{x},t)$ is the same as in Eq.~\eqref{correlationFtheta} with $f(t-s)$ substituting $\delta(t-s)$.
As already derived in Ref.~\cite{carlesso2018colored}, the colored density noise spectrum can be defined in terms of the white one, which is shown in Eq.~\eqref{DNS}:
\begin{equation}
    S^\text{\tiny C}_\text{\tiny CSL}(\omega)=\tilde f(\omega) \times S_\text{\tiny CSL}(\omega).
\end{equation}
where $\tilde f(\omega)$ is the Fourier transform of $f(t)$.
We consider a exponential correlation function $f$, which is characteristic of many physical processes, as already done Ref.~\cite{bassi2009non}:
\begin{equation}
    f(t-s)=\frac{\Omega_\text{\tiny C}}{2}e^{-\Omega_\text{\tiny C}|t-s|},
\end{equation}
with correlation time $\Omega_\text{\tiny C}^{-1}$; by doing this we introduce a cutoff in the frequency domain. Correspondingly we obtain the following DNS:
\begin{equation}
    S^\text{\tiny C}_\text{\tiny CSL}(\omega)=\frac{\Omega^2_\text{\tiny C}}{\Omega^2_\text{\tiny C}+\omega^2} S_\text{\tiny CSL}(\omega).
\end{equation}
As long as $\Omega_\text{\tiny C}\gg \omega\simeq 10^{-2}\,\mathrm{s^{-1}}$, $S^\text{\tiny C}_\text{\tiny CSL}(\omega)\simeq S_\text{\tiny CSL}(\omega)$, meaning that the results derived in the main text are not affected by the cutoff.
For comparison, the bounds on the CSL parameters coming from the spontaneous radiation emission set in Ref.~\cite{arnquist2022search} for the 
\textsc{Majorana Demonstrator}
remain valid only for values of the cutoff $\Omega_\text{\tiny C}\gg 10^{19}\;\mathrm{s^{-1}}$.
As discussed in Ref.~\cite{carlesso2018colored}, a reasonable value for the cutoff frequency is $\Omega_\text{\tiny C}\simeq 10^{12}\,$s$^{-1}$, which leaves unaffected the bound derived in this work, but suppresses the bound set by radiation emission experiments. 
\section{Theoretical excluded region}\label{appD}

\ario{We comment on the theoretical lower bound on the CSL parameters.
The bound represented in Fig.~\ref{fig:BOUND} was obtained by considering a graphene disk with diameter of $20\,\mathrm{\mu m}$ (about the smallest possible size detectable by human eye) and requiring it to collapse in less than $0.01\,\mathrm{s}$ (about the time resolution of human eye) \cite{torovs2017colored}. The area is coloured with a gradient since there is some degree of subjectivity in choosing the system’s size and the time within which the superposition must collapse.
For example, another comparable bound was found by requiring that a carbon sphere with the diameter of $4000\,\mathrm{\AA}$ must collapse in less than $0.01\,\mathrm{s}$ \cite{collett1995constraint}. Finally, a much weaker bound was proposed in Ref.~\cite{feldmann2012parameter}, by requiring the collapse of ink molecules corresponding to a digit in a printout in less than $0.5\,\mathrm{s}$ (for a graphical representation, see Fig.~4 
 of \cite{donadi2021novel}).}

\bibliography{main.bib}{}
\bibliographystyle{apsrev4-1}

\end{document}